\documentclass[journal=jctcce,manuscript=article]{achemso}
\usepackage{amssymb,amsmath,amsfonts,multicol,multirow,longtable,array,mathpazo}
\usepackage{lscape}
\usepackage[version=3]{mhchem} % Formula subscripts using \ce{}
\usepackage{appendix}
\usepackage{soul} %text highlighting
\usepackage[usenames]{color}
\usepackage{makecell}
\usepackage{enumerate}
\usepackage{appendix}
\usepackage{xr}
\usepackage[T1]{fontenc} % Use modern font encodings
\usepackage{booktabs}
\usepackage{tikz}
\usepackage{threeparttable}
\usepackage{graphicx}
\usepackage{subfigure}
\usepackage{colortbl}
\usepackage{framed}
\usepackage{verbatim}
\usepackage{amsbsy}
\usepackage{bm}% bold math
\usepackage{bbm}
\usepackage[normalem]{ulem}
\usepackage{float}
\usepackage[linesnumbered,boxed]{algorithm2e}
\usepackage{algorithm2e}
\usepackage{amsmath}
\usepackage{diagbox}
\usepackage{changepage}
\usepackage{threeparttable}

\mciteErrorOnUnknownfalse
\newcounter{xscheme}

\newfloat{Algorithm}{htbp}{alg}
\floatname{Algorithm}{Algorithm}

\newcounter{exe}[figure]
\newcommand{\iexe}{\refstepcounter{exe}\the\value{exe}:}

\setkeys{acs}{maxauthors = 0} % references of many authors

\author{Yang Guo}
\affiliation{Institute of Frontier Chemistry, School of Chemistry and Chemical Engineering, Shandong University, Qingdao, Shandong 266237, China}
\email{yang.guo@sdu.edu.cn}

\author{Achintya Kumar Dutta}
\affiliation{Department of Chemistry, Indian Institute of Technology Bombay, Powai, Mumbai 400076, India}
\title{A Perturbative Super-CI Approach for orbital optimization in Two-Component relativistic CASSCF}
\setkeys{acs}{articletitle=true}

\begin{document}
%\twocolumn[\begin{@twocolumnfalse}
%\end{@twocolumnfalse}]

\newpage

\begin{abstract}

In this work, we develop a new orbital optimization approach, perturbative Super-CI (Super-CIPT), for the two-component complete active space self-consistent field (2C-CASSCF) method. By variationally optimizing spinor orbitals and consistently incorporating spin--orbit coupling (SOC) at the orbital level, the 2C-CASSCF method enables a  simultaneous treatment of relativistic effects and static correlation. The Super-CIPT approach demonstrates robust convergence behavior and is applicable to systems under strong SOC. The inclusion of Gaunt or Breit term via the atomic mean field approximation yields the most accurate results, with errors dropping below 2\% for halogens. We systematically assess the performance of 2C-CASSCF on spin-orbit splittings (SOSs) of selected $p$-block elements. Results show that 2C-CASSCF outperforms conventional one-component (1C) CASSCF. This work establishes 2C-CASSCF with Super-CIPT as a reliable and efficient approach for multireference relativistic quantum chemistry.

\end{abstract}

\maketitle

\clearpage
\newpage

\section{Introduction}
Accurate electronic structure calculations for heavy-element systems demand a balanced treatment of electron correlation and relativistic effects. As the nuclear charge increases, relativistic contributions such as scalar relativistic effects, spin–orbit coupling, and picture-change effects significantly influence molecular energies and properties\cite{RQC_BOOK_Dyall}. Consequently, multi-reference methods formulated within a relativistic framework are indispensable for describing systems with near-degeneracy, open-shell character, or strong spin–orbit interactions, which commonly arise in heavy atoms and their compounds. Relativistic quantum chemical methods can be broadly classified into two categories: those based on scalar orbitals and those based on spinor orbitals. Spinor-based approaches are further divided into four-component (4C) \cite{4C-RASCI1993,4C-GASCI2003,4C-MCSCF1996,4C-CI-MCSCF2006,4C-MCSCF2008,
4C-CASPT22008,4C-CASSCF2015,4C-CASSCF2018,4C-icMRCI2015,4C-MR2018,4C-DMRG2014,4C-DMRG2018,
4C-DMRG2020,MPSSI2021,4c-FCIQMC,4C-MRMP21999,4C-CASPT22006} and two-component (2C)\cite{2C-CASSCF1996,2C-CI2001,2C-CASSCF2003,2C-CASSCF2013,LixiaosongX2CCASSCF,
2C-MRPT22014,LiXiaosongX2C-MRPT22022,2C-MRCI2020,X2C-DMRG-Li-2022} formalisms. The 4C methods, rooted in the full relativistic equation, treat both the large and small components of the spinors explicitly and account for relativistic effects, including spin–orbit coupling (SOC), to all orders, enabling a rigorous description of electronic structure in heavy-element systems. In contrast, 2C methods employ unitary transformations to decouple the large and small components at the Hamiltonian level, resulting in a computationally efficient formalism that retains the accuracy of the 4C framework. Among the various two-component methods available, the exact two-component (X2C) approach is the most popular.\cite{X2C2005,X2C2009,NESC} Over the past decades, various single-reference and multireference 4C and 2C quantum chemistry methods have been developed and successfully applied to systems involving heavy elements.\cite{LiuPerspective2020}

Multiconfigurational self-consistent field (MCSCF) methods are essential for accurately describing electronic systems where static correlation plays a dominant role, such as in bond dissociation, radicals, and excited states with strong configurational mixing. Among these, the complete active space self-consistent field (CASSCF) method stands out by defining an active space in which all possible Slater determinants (SDs) or configuration state functions (CSFs) are included in the wave function expansion.\cite{FORS1978,CASSCF,WernerRev1987,CASSCFRev1987,MCSCFRev1987,MCSCFRev1998,MCSCFrev2012} The orbitals in CASSCF are variationally optimized in a self-consistent manner, allowing for orbital relaxation that is critical in strongly correlated systems. However, in molecules containing heavy atoms, relativistic effects, particularly SOC effect, become significant and must be incorporated to achieve reliable predictions of spectroscopic properties and electronic structures.

The extension of MCSCF to relativistic regimes has led to three main classes of methods, those based on scalar orbitals (denoted as 1C), as well as those based on 2C and 4C spinor orbitals. In scalar orbital frameworks, scalar relativistic effects are typically included via spin-free one-body Hamiltonians (e.g., spin-free X2C).\cite{X2CSOC1,X2CSOC2} To further include SOC effect, various two-step and one-step scalar CASSCF methods are reported. Two-step (or state-interaction) methods treat SOC as a perturbation acting on converged CASSCF wave functions, assuming additivity between electron correlation and SOC.\cite{CASPT2-SOC2004} This assumption, however, breaks down in cases where SOC and correlation are strongly entangled, and may lead to inaccurate descriptions of fine-structure splittings and spin state ordering. In contrast, one-step approaches incorporate SOC variationally at the mean-field level using spin-dependent SDs or CSFs, enabling a consistent treatment of SOC and electron correlation, along with SOC-induced orbital relaxation.\cite{SOCASSCF2013} Despite their advantages, both one- and two-step methods are typically limited to active spaces of up to 20 scalar orbitals due to exponential scaling. To extend their applicability, selected configuration interaction (sCI) techniques have been integrated into one-step frameworks by Sharma\cite{HBCISOC} and Liu\cite{SOiCI,SOiCISCF} independently. 

However, scalar-orbital-based methods are fundamentally insufficient for systems that exhibit both strong relativistic effects and strong electron correlation, particularly those incorporating elements from the sixth and seventh periods. In such cases, a spinor-based MCSCF formulation is required for a more accurate and balanced description. This has motivated the development of 2C and 4C MCSCF methods, which operate directly in the spinor orbital basis and includes a variational treatment of the spin-orbit coupling. The first 4C MCSCF method was reported by Jensen and coworkers in 1997,\cite{4C-MCSCF1996} followed by systematic developments of 4C and 2C general active space SCF (GASSCF) methods by Fleig et al.\cite{2C-CASSCF1996,2C-CI2001,4C-GASCI2003,4C-CI-MCSCF2006,4C-MCSCF2008,Fleig2012} Lee and coworkers also developed a 2C-CASSCF approach based on relativistic effective core potentials.\cite{2C-CASSCF2003,2C-CASSCF2013} In the past decade, Shiozaki and coworkers significantly improved the convergence and scalability of 4C-CASSCF implementations.\cite{4C-CASSCF2015,4C-CASSCF2018} More recently, Li and coworkers have also contributed to the development of both 2C- and 4C-CASSCF methods.\cite{LixiaosongX2CCASSCF,DMCSCF,X2C-DMRG-Li-2022} Despite these advances, the computational cost of 2C- and 4C-CASSCF methods scales rapidly with the number of spinor orbitals in the active space, limiting practical applications to limited spinor orbitals in most cases. To address this bottleneck, density matrix renormalization group \cite{4C-DMRG2018,X2C-DMRG-Li-2022} and selected CI\cite{2C/4C-SHCI} based approaches have been combined with 2C and 4C frameworks for treating larger active spaces.

In this work, we present a 2C-CASSCF method that variationally optimizes spinor orbitals, enabling a simultaneous and self-consistent treatment of static correlation and relativistic effects. The spinor orbitals are optimized using the perturbative Super-CI (Super-CIPT) approach developed by Kollmar and coworkers.\cite{Super-CIPT} As an efficient first-order optimization approach, Super-CIPT significantly reduces computational cost relative to conventional second-order orbital optimization methods, particularly in the evaluation of two-electron integrals.  
%\textcolor{red}{The accuracy of the method is assessed using spin–orbit splittings (SOSs) of $p$-block elements. The convergence behavior of Super-CIPT is examined, and the method is applied to compute the low-lying excited states of HI and HAt molecule, demonstrating its capability in capturing static correlation and relativistic effects in a unified framework(\textbf{may be we can omit this lines})}.

\section{Super-CIPT for 2C-CASSCF}\label{Theory}

Throughout this work, core, active, virtual, and arbitrary spinor orbital indices are denoted by ${i,j,k,l,\cdots}$, ${t,u,v,w,\cdots}$, ${a,b,c,d,\cdots}$, and ${p,q,r,s,\cdots}$, respectively. A spinor orbital obtained from 2C HF or CASSCF calculations is represented as $|\phi_p\rangle$. The CASCI wave function constructed from core and active spinor orbitals is denoted by $|\Psi^{\text{CI}}\rangle$.

The electronic 2C Hamiltonian can be generally expressed as  
\begin{eqnarray} 
\hat{H} = \sum_{pq} h_{pq} a^{\dagger}_{p} a_{q} + \frac{1}{2} \sum_{pqrs} g_{pqrs} a^{\dagger}_{p} a^{\dagger}_{r} a_{s} a_{q}, 
\end{eqnarray}  
irrespective of the approximations employed for the one- and two-electron integrals, $h_{pq}$ and $g_{pqrs}$. By performing a CASCI calculation with the 2C Hamiltonian, the final CASCI energy could be written as  
\begin{eqnarray} 
E = \langle \Psi^{\text{CI}} | \hat{H} | \Psi^{\text{CI}} \rangle = \sum_{pq} h_{pq} \gamma_{pq} + \frac{1}{2} \sum_{pqrs} g_{pqrs} \gamma_{pqrs} 
\end{eqnarray}  
where  
\begin{eqnarray} 
\gamma_{pq} &=& \langle \Psi^{\text{CI}} | a^{\dagger}_p a_q | \Psi^{\text{CI}} \rangle \qquad (\tilde{\gamma}_{pq} = \mathbf{I} - \gamma_{pq}), \\ 
\gamma_{pqrs} &=& \langle \Psi^{\text{CI}} | a^{\dagger}_p a^{\dagger}_r a_s a_q | \Psi^{\text{CI}} \rangle. 
\end{eqnarray}
In 2C-CASSCF, the spinor orbitals are optimized until convergence is achieved. To facilitate this optimization, the orbital rotation operator $ \exp(\hat{\kappa}) $ is applied to the wave function, leading to the energy expression,
\begin{equation}
E = \left\langle \Psi^{\text{CI}} \left| \exp(-\hat{\kappa}) \hat{H} \exp(\hat{\kappa}) \right| \Psi^{\text{CI}} \right\rangle,
\quad
(\hat{\kappa} = \sum_{pq} \kappa_{pq} a^{\dagger}_p a_q) \quad.
\end{equation}
At the converged 2C-CASSCF wave function, the orbital gradients, defined as the partial derivatives of the energy $E$ with respect to the orbital rotation parameters $\kappa_{pq}$, vanish,
\begin{eqnarray}
G_{pq} = \frac{\partial E}{\partial \kappa_{pq}} = \frac{\partial}{\partial \kappa_{pq}} \left\langle \Psi^{\text{CI}} \left| \exp(-\hat{\kappa}) \hat{H} \exp(\hat{\kappa}) \right| \Psi^{\text{CI}} \right\rangle = 0 \quad.
\end{eqnarray}
In 2C-CASSCF, the rotation parameters are determined iteratively until convergence achieved.

In Super-CIPT,\cite{Super-CIPT} the parameter $\kappa_{pq}$ is determined according to the following equation:
\begin{eqnarray}
\sum_{pq}\kappa_{pq} \left\langle \Psi^{\text{CI}}\middle|a^{\dagger}_r a_s\hat{H}^{Dyall}a^{\dagger}_p a_q\middle|\Psi^{\text{CI}}\right\rangle + \left\langle \Psi^{\text{CI}}\middle|\hat{H}a^{\dagger}_r a_s\middle|\Psi^{\text{CI}}\right\rangle = 0 \quad.\label{SuperCIPT0}
\end{eqnarray}
The Dyall Hamiltonian in the spinor basis can be defined as 
\begin{eqnarray}
\hat{H}^{Dyall} &=& \sum_{ij}f_{ij}a^{\dagger}_j a_i+\sum_{ab}f_{ab}a^{\dagger}_b a_a+\sum_{tu}h_{tu}a^{\dagger}_t a_u+\frac{1}{2}\sum_{tuvw} g_{tuvw} a^{\dagger}_t a^{\dagger}_v a_w a_u + C, \label{Dyall}\\
f_{pq} &=& f^{occ}_{pq} + f^{act}_{pq} = h_{pq}+\sum_{i}(g_{pqii}-g_{piiq}) +\sum_{tu}(g_{pqtu}-g_{ptuq})\gamma_{tu}.
\end{eqnarray}
As in the scalar case, the $C$ in Eq. \ref{Dyall} ensures that $ |\Psi^{\text{CI}}\rangle $ is an eigenfunction of the Dyall Hamiltonian, with the eigenvalue $E$, 
\begin{eqnarray}
\hat{H}^{Dyall} |\Psi^{\text{CI}}\rangle = E |\Psi^{\text{CI}}\rangle. \label{Dyall_eig}
\end{eqnarray}

With the above relation, Eq. \ref{SuperCIPT0} can be simplified as,
\begin{eqnarray}
\sum_{pq}\kappa_{pq} \left\langle \Psi^{\text{CI}}|a^{\dagger}_r a_s[\hat{H}^{Dyall},a^{\dagger}_p a_q]|\Psi^{\text{CI}}\right\rangle + \sum_{pq}\kappa_{pq}E\left\langle \Psi^{\text{CI}}|a^{\dagger}_r a_s a^{\dagger}_p a_q|\Psi^{\text{CI}}\right\rangle
=-\left\langle \Psi^{\text{CI}}|\hat{H}a^{\dagger}_r a_s|\Psi^{\text{CI}}\right\rangle \label{SuperCIPT1}
\end{eqnarray}
The spinor orbitals in 2C-CASSCF can be partitioned into three spaces as well, the core, active, and virtual spaces. The rotations within the core, active, and virtual spaces are redundant. Thus, the orbital rotation parameters can be classified into three types,
\begin{eqnarray}
\hat{\kappa} = \sum_{ai} \kappa_{ai} a^{\dagger}_a a_i + \sum_{ti} \kappa_{ti} a^{\dagger}_t a_i + \sum_{at} \kappa_{at} a^{\dagger}_a a_t.
\label{Kappa}
\end{eqnarray}
Thus, Eq. \ref{SuperCIPT1} can be decomposed into three independent sets of equations,
\begin{eqnarray}
\kappa_{ai} \left\langle \Psi^{\text{CI}}|a^{\dagger}_j a_b[\hat{H}^{Dyall}, a^{\dagger}_a a_i]|\Psi^{\text{CI}}\right\rangle + \kappa_{ai}E \left\langle \Psi^{\text{CI}}|a^{\dagger}_j a_b a^{\dagger}_a a_i|\Psi^{\text{CI}}\right\rangle = G_{bj} \quad,\\ 
\kappa_{ti} \left\langle \Psi^{\text{CI}}|a^{\dagger}_j a_u[\hat{H}^{Dyall},a^{\dagger}_t a_i]|\Psi^{\text{CI}}\right\rangle +\kappa_{ti}E \left\langle \Psi^{\text{CI}}|a^{\dagger}_j a_u a^{\dagger}_t a_i|\Psi^{\text{CI}}\right\rangle = G_{uj} \quad, \\ 
\kappa_{at} \left\langle \Psi^{\text{CI}}|a^{\dagger}_u a_b[\hat{H}^{Dyall}, a^{\dagger}_a a_t]|\Psi^{\text{CI}}\right\rangle + \kappa_{at}E \left\langle \Psi^{\text{CI}}|a^{\dagger}_u a_b a^{\dagger}_a a_t|\Psi^{\text{CI}}\right\rangle = G_{bu} \quad, 
\label{SuperCIPT2}
\end{eqnarray}
where the orbital gradients $G_{pq}$ are evaluated by,
\begin{eqnarray}
G_{pq}=\left\langle \Psi^{\text{CI}}|[ \hat{H},a^{\dagger}_p a_q ]|\Psi^{\text{CI}}\right\rangle 
&=&\left\langle \Psi^{\text{CI}}|a^{\dagger}_p a_q \hat{H}|\Psi^{\text{CI}}\right\rangle - \left\langle \Psi^{\text{CI}}|\hat{H} a^{\dagger}_p a_q|\Psi^{\text{CI}}\right\rangle \nonumber \\
&=& - \langle \Psi^{\text{CI}}|\hat{H} a^{\dagger}_p a_q|\Psi^{\text{CI}}\rangle \quad.
\end{eqnarray}
Note that $p$ and $q$ belong to different subspaces.
By defining 
\begin{eqnarray}
\tilde{K} \tilde{C}_{\mu}= \tilde{\epsilon}_{\mu}\tilde{\gamma} \tilde{C}_{\mu}\\ 
K C_{\lambda}= \epsilon_{\lambda}\gamma C_{\lambda} 
\end{eqnarray}
where
\begin{eqnarray}
K_{tu} &=& \langle \Psi^{CI}|a^{\dagger}_u [\hat{H}^{Dyall}, a_t]|\Psi^{CI}\rangle = -\sum_{t'}f^{occ}_{tt'}\gamma_{t'u}- \sum_{t'u'v'}g_{tt'u'v'}\gamma_{ut'u'v'}\\
\tilde{K}_{ut} &=& \langle \Psi^{CI}| a_u[\hat{H}^{Dyall},a^{\dagger}_t]|\Psi^{CI}\rangle=f_{tu}+K_{tu}\\ 
\end{eqnarray}
The final rotational parameters are determined as follows, 
\begin{eqnarray}
\kappa_{ai} &=& \frac{G_{ai}}{f_{ii}-f_{aa}}\\
\kappa_{ti} &=& \sum_{\mu}\frac{\sum_{v}\tilde{C}_{\mu v}G_{vi}}{f_{ii}-\sum_{u}\tilde{\epsilon}_{\mu}} \sum_{u}\tilde{C}_{\mu u}\tilde{\gamma}_{ut}\\
\kappa_{at} &=& \sum_{\lambda}\frac{\sum_{v}C_{\lambda v}G_{av}}{-\epsilon_{\lambda}-f_{aa}} \sum_{u}C_{\lambda u}\gamma_{ut}
\end{eqnarray}

Using the anti-Hermitian matrix $\kappa_{pq}$, the spinor orbitals are updated as  
\begin{eqnarray} 
|\phi^{new}_i\rangle = \sum_k |\phi^{old}_k\rangle U_{ki}, \quad \text{where} \quad U = \exp(\hat{\kappa}) \quad. 
\end{eqnarray}  
The optimization procedure terminates upon convergence, which is reached when the norm of the orbital gradient approaches zero.

\section{Computational details}
The 2C-CASSCF approach has been implemented in a development version of the BAGH package\cite{BAGH} interfaced to PySCF \cite{pyscf1,pyscf2} and socutils\cite{wangXubwaSocutils2025}.  For the calculation of SOS of $p$-block elements, four different active spaces are used, consisting of $np$, $nsnp$,  $np(n+1)p$, and $nsnp(n+1)s(n+1)p$ electrons and orbitals, respectively. To be consistent with the CASSCF with scalar orbitals, the active space, consisting of $m$ active electrons and $n$ pairs of active spinor orbitals is denoted as 2C-CAS($m$,$n$). Note that $n$ in X2C-CASSCF contains 2$n$ spinor orbitals. Thus, the four active spaces for $p$-block elements are denoted as 2C-CAS($m$,3), 2C-CAS($m$+2,4), 2C-CAS($m$,6), and 2C-CAS($m$+2,8). To distinguish from 2C-CASSCF, the 1C-CASSCF calculations performed by the state-interaction CASSCF method are denoted as 1C-CAS($m$,$n$). In 1C-CAS calculations, SOC effects are considered via the state interaction method. The first-order Douglas-Kroll-Hess type of effective one-body spin-orbit (so-DKH1) operator is used as the SOC Hamiltonian.\cite{DKH1} The X2C scalar relativistic effect is also considered.\cite{X2CSOC1} The Dyall's uncontracted triple-$\zeta$ basis sets, dyallv3z,\cite{dyalltz} are used in all calculations. 

In 2C-CASSCF, different Hamiltonians could be employed. The simplest variant is X2C1e (exact two-component with only one-electron terms). Starting from the 4C Hamiltonian, the small component could be eliminated via a unitary transformation, leading to a two-component Hamiltonian,
\begin{equation}
H^{\text{X2C1e}} = \sum_{pq} h_{pq}^{\text{X2C}} a_p^\dagger a_q + \frac{1}{2} \sum_{pqrs} g_{pqrs}^{\text{NR}} a_p^\dagger a_q^\dagger a_s a_r,
\end{equation}
where $h^{\text{X2C}}$ is the one-electron X2C matrix obtained by block-diagonalizing the Dirac Hamiltonian using the X2C transformation matrices $X$ and $R$:
\begin{equation}
h^{\text{X2C}} = R^\dagger \left( V + T + X^\dagger T X + \frac{1}{4} X^\dagger W X \right) R + T X R \quad (W = (\boldsymbol{\sigma} \cdot \mathbf{p}) V (\boldsymbol{\sigma} \cdot \mathbf{p})),
\end{equation}
with $V$ the nuclear potential and $T$ the kinetic operator. In $W$, $\boldsymbol{\sigma}$ represents the Pauli matrices, and $\mathbf{p}$ represents the momentum operator. The two-electron integrals $g^{\text{NR}}$ are taken in their nonrelativistic (NR) form, neglecting picture-change effects. To further consider the two-body operators, the exact two-component Hamiltonian with atomic mean-field (X2CAMF) approximation to the Hamiltonian could be used \cite{Cheng2C-CC2018,X2CAMF2022,X2CAMF_rev},
\begin{eqnarray}
\hat{H}^{\text{X2CAMF}} = \sum_{pq} (h_{pq}^{\text{X2C}} + g_{pq}^{4c, \text{AMF}}) a^\dagger_p a_q + \frac{1}{2} \sum_{pqrs} g_{pqrs}^{\text{NR}} a^\dagger_p a^\dagger_r a_s a_q,
\end{eqnarray}
where
\begin{eqnarray}
\sum_{pq} g_{pq}^{4c, \text{AMF}} a^\dagger_p a_q = \sum_A \sum_i \sum_{pq} n_{i_A} (g_{pqi_Ai_A}^{4c, \text{SD}} - g_{pi_Ai_Aq}^{4c, \text{SD}}) a^\dagger_p a_q\quad. \label{AMF}
\end{eqnarray}
In Eq. (\ref{AMF}), $g^{\text{4c,SD}}$ represent spin-dependent (SD) two-electron integrals. In the present work, three different X2CAMF Hamiltonians are considered, depending on the 4C relativistic Hamiltonian, the Dirac-Coulomb (DC), Dirac-Coulomb-Gaunt (DCG), and Dirac-Coulomb-Breit (DCB) Hamiltonians. 

In this work, the 1C-CASSCF calculations are performed using ORCA6.0.\cite{ORCA,ORCA5,ORCAJCP} The 4C-CASSCF calculations are performed using BAGEL.\cite{BAGEL} The auxiliary basis sets required by BAGEL are generated by AutoAux\cite{AutoAux} utility of ORCA.

\section{Results and Discussion}\label{Results}

\subsection{Hamiltonian}

The spin--orbit splitting (SOS) of the $^2P$ configuration of halogen atoms (Cl, Br, I, At) is calculated using 2C-CASSCF with the CAS(7,4) active space, employing various relativistic Hamiltonians: X2C1e, X2CAMF(DC), X2CAMF(DCG), and X2CAMF(DCB). As shown in Table \ref{Hamiltonian}, the simplest X2C1e approximation yields large absolute errors, particularly for heavier elements, 1438.9~cm$^{-1}$ for At, corresponding to a relative error of 6.15\%. This significant deviation indicates the inaccuracy of X2C1e in capturing SOC effects due to its neglect of two-electron relativistic contributions.

In contrast, the X2CAMF(DC) Hamiltonian, which includes the AMF approximation to the two-electron Coulomb term, dramatically improves the accuracy. For Br, I, and At, the relative errors are reduced to below 3\%, with only Cl showing a slightly larger discrepancy (5.32\%). Further inclusion of the Gaunt (DCG) or Breit (DCB) corrections within the X2CAMF framework leads to even better agreement with experimental values, reducing relative errors to below 2\% across all halogens. The best 2C results are achieved for bromine, where the error drops to merely 0.06\% with the X2CAMF(DCG) or X2CAMF(DCB) Hamiltonian.

For comparison, 4C-CASSCF results of halogen atoms are also reported. For Cl and Br, the 2C and 4C approaches yield nearly identical SOS values when using the same type of Hamiltonian, with absolute differences less than 1.0 cm$^{-1}$. This excellent agreement confirms that the X2CAMF approximation precisely recovers the 4C results for lighter elements. As atomic number increases, small deviations between 2C and 4C emerge, due to the growing importance of relativistic effects. Nevertheless, even for astatine, the difference in relative error between 2C and 4C is less than 0.12\%, indicating that the X2CAMF approximation remains highly accurate even for the heaviest halogens.

Between the two extended Hamiltonians, X2CAMF(DCG) yields slightly smaller errors than X2CAMF(DCB), but the error may be due to fortuitous error cancellation and the differences are marginal. Given the comparable performance and similar computational cost, the X2CAMF(DCB) Hamiltonian is adopted for subsequent calculations in this work.

\begin{table}[htbp]
\centering
\footnotesize
\begin{threeparttable}
\caption{The SOS (in cm$^{-1}$) of Cl, Br, I, and At calculated using 2C- and 4C-CASSCF methods with various Hamiltonians at CAS(7,4)/dyallv3z level. Absolute errors with respect to experimental values (Expt.) are listed.}
\label{Hamiltonian}
\setlength{\tabcolsep}{0pt}
\begin{tabular*}{\textwidth}{@{\extracolsep{\fill}} crrrrrrrr @{}}
\hline
\multirow{2}{*}{\textbf{Atom}} & 
\multicolumn{1}{c}{\textbf{X2C1e}} & 
\multicolumn{2}{c}{\textbf{DC}} & 
\multicolumn{2}{c}{\textbf{DCG}} & 
\multicolumn{2}{c}{\textbf{DCB}} & 
\multicolumn{1}{c}{\textbf{Expt.}} \\
&   \multicolumn{1}{c}{2C}&
  \multicolumn{1}{c}{2C} & \multicolumn{1}{c}{4C} & 
  \multicolumn{1}{c}{2C} & \multicolumn{1}{c}{4C} & 
  \multicolumn{1}{c}{2C} & \multicolumn{1}{c}{4C} & (cm$^{-1}$)\\
\hline
Cl	&  203.1(23.01\%) 	& 46.9(5.32\%) 	& 46.1(5.22\%) 	&  7.3(0.83\%)  &  7.1(0.80\%) 	&  7.4(0.84\%) 	&7.0(0.80\%) 	 &882.4   \\
Br	&  385.4(10.46\%) 	& 73.5(2.00\%) 	& 73.6(2.00\%) 	&  2.0(0.06\%)  &  2.7(0.07\%) 	&  2.2(0.06\%) 	&3.1(0.09\%) 	 &3685.2  \\
I	  &  599.4( 7.88\%) 	&160.9(2.12\%) 	&163.3(2.15\%) 	& 61.3(0.81\%)  & 64.1(0.84\%) 	& 61.1(0.80\%) & 64.7(0.85\%)  &7603.1  \\         
At	& 1438.9( 6.15\%) 	&634.1(2.71\%) 	&654.6(2.80\%) 	&389.8(1.67\%)  &415.7(1.78\%)\tnote{a}  &392.8(1.68\%) &406.1(1.74\%)\tnote{a} &23393.8\tnote{b} \\ 
\hline
\end{tabular*}
\label{Hamiltonian}
\begin{tablenotes}
\small
\item[a] Not fully converged.
\item[b] mmf-X2C-FS-CCSD result~\cite{sf-X2C-EOM-SOC2017}.
\end{tablenotes}
\end{threeparttable}
\end{table}

\subsection{Convergence trend}

\begin{figure}[!htp]
        \centering
                \begin{tabular}{c}
                    \includegraphics[width=1.0\textwidth]{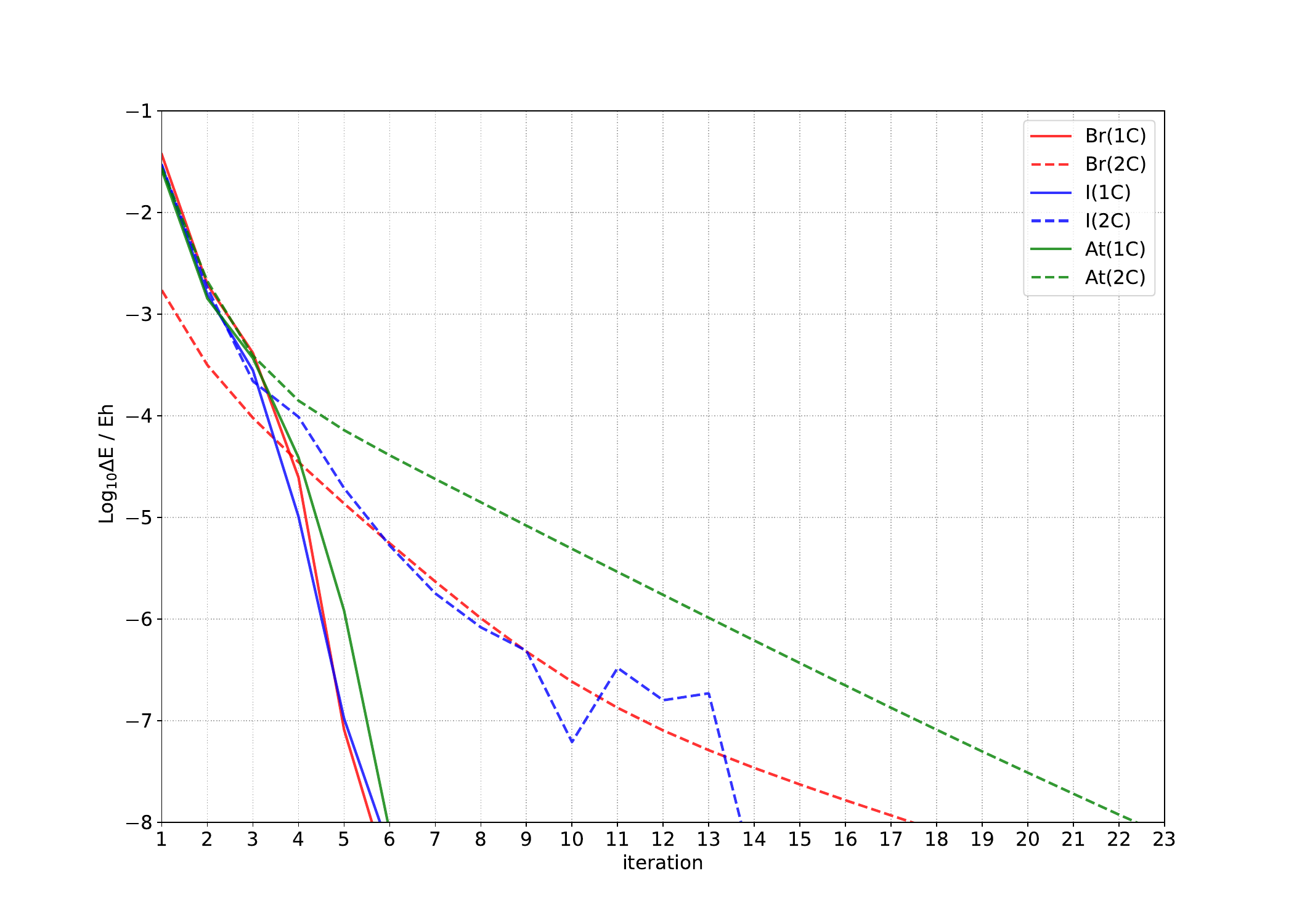}\\
                \end{tabular}
                \caption{Convergence patterns of 1C- and 2C-CAS(7,4) calculations of Br, I, and At.}
        \label{conv}
\end{figure}

The convergence trends of the Super-CIPT approach for the 2C-CASSCF calculations are accessed using the Br, I, and At atoms. Six spinor states of the $^2P$ configuration are computed with averaged weight. For comparison, the 1C-CASSCF for the three degenerate spin-free states are computed with the Super-CIPT approach as well.\cite{Super-CIPT} The convergence behavior of 1C- and 2C-CASSCF calculations for Br, I, and At are given in Fig. \ref{conv}, based on the energy difference between successive iterations. In the 1C-CASSCF calculations, the Super-CIPT approach converged much faster than those in 2C-CASSCF. For all the three 1C-CASSCF calculations, the orbital optimization converged within 6 iterations. In contrast, the Super-CIPT approach for the At element needs 23 iterations, which converges slower than the calculations of bromine and iodine. The slower convergence of 2C-CASSCF should be due to the increased complexity from relativistic effects, especially the SOC effect. The 2C-CASSCF calculations of Br and At converged monotonically. In contrast, for iodine, the energy convergence exhibits a small oscillation between iterations 10 and 13, fluctuating between $10^{-6}$ and $10^{-8}$ Hartree, before resuming monotonic descent. This transient behavior reflects the complex interplay between orbital relaxation and SOC during the orbital optimization process. 
%It should be noted that, initially, the 1C-CASSCF calculation for iodine converges more rapidly than that for bromine; however, its convergence rate decreases after the fifth iteration.

\subsection{SOS of $p$-block elements}

\begin{figure}[!htp]
        \centering
                \begin{tabular}{c}
\includegraphics[width=1.0\textwidth]{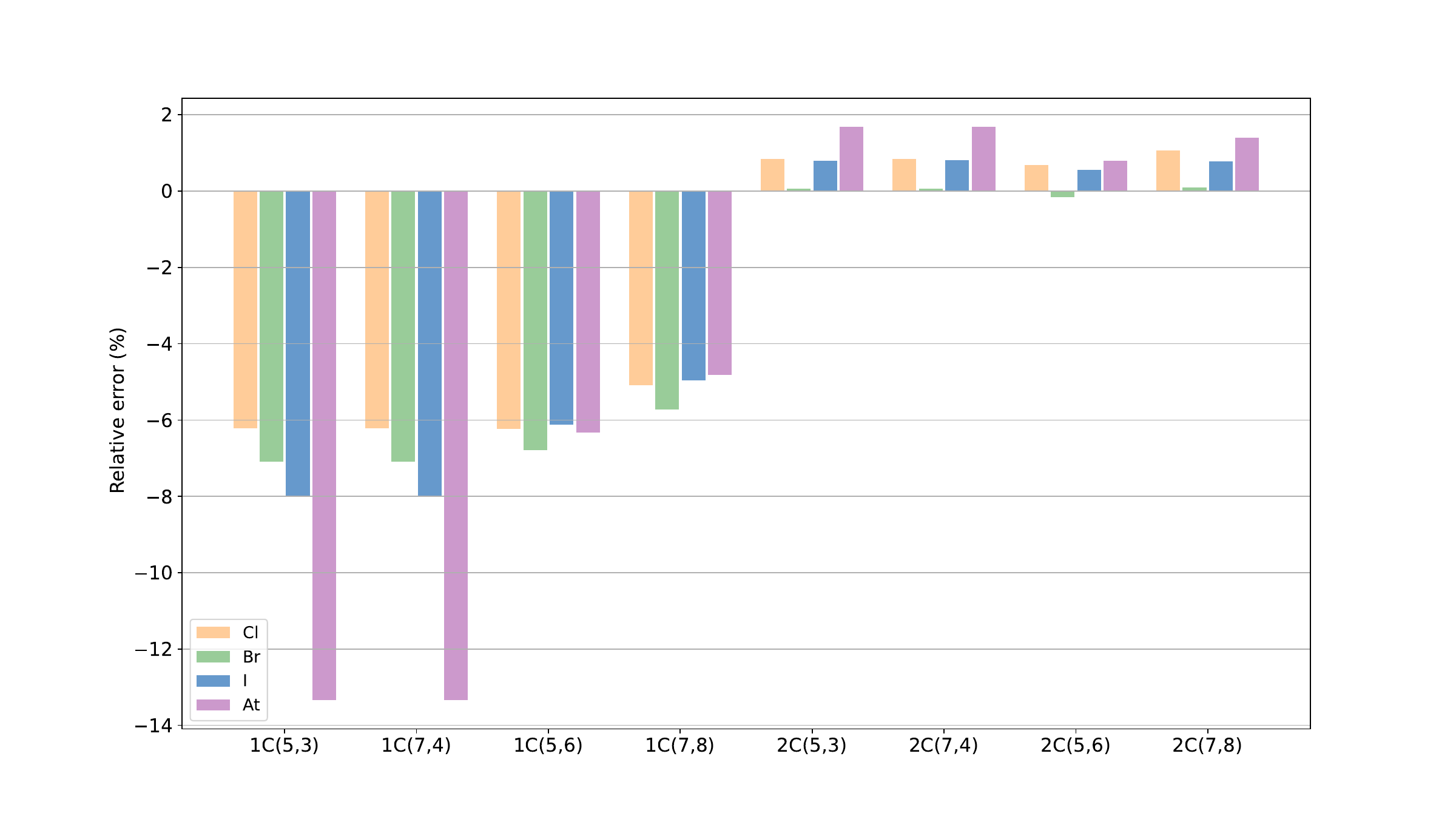}\\
                \end{tabular}
                \caption{The relative errors of SOSs of $^2P$ configuration of Cl, Br, I, and At computed by 1C- and 2C-CASSCF with four different active spaces.}
        \label{Rel_7}
\end{figure}

The SOSs of $^2P$ configuration of halogen elements are computed using 1C-CASSCF and 2C-CASSCF with four different active spaces (Figure. \ref{Rel_7}). The 1C-CASSCF method systematically underestimates the SOS across all atoms and active spaces, with relative errors increasing significantly down the group, from -6.22\% for Cl to -13.34\% for At with 1C-CAS(5,3), indicating an inadequate treatment of SOC effects in the scalar reference. In contrast, the 2C-CASSCF method achieves higher accuracy, with relative errors reduced to below 1.7\% for all cases, and as low as 0.06\% for Br. Notably, the 2C-CASSCF results show much weaker dependence on the choice of active space, whereas the 1C approach exhibits larger variations with different active spaces, particularly for heavier elements. This highlights the importance of using 2C Hamiltonian and wave functions. The satisfactory result from 2C-CASSCF underscores its reliability for high-precision predictions of SOSs in halogen elements.

\begin{figure}[!htp]
        \centering
                \begin{tabular}{c}
                    \includegraphics[width=1.0\textwidth]{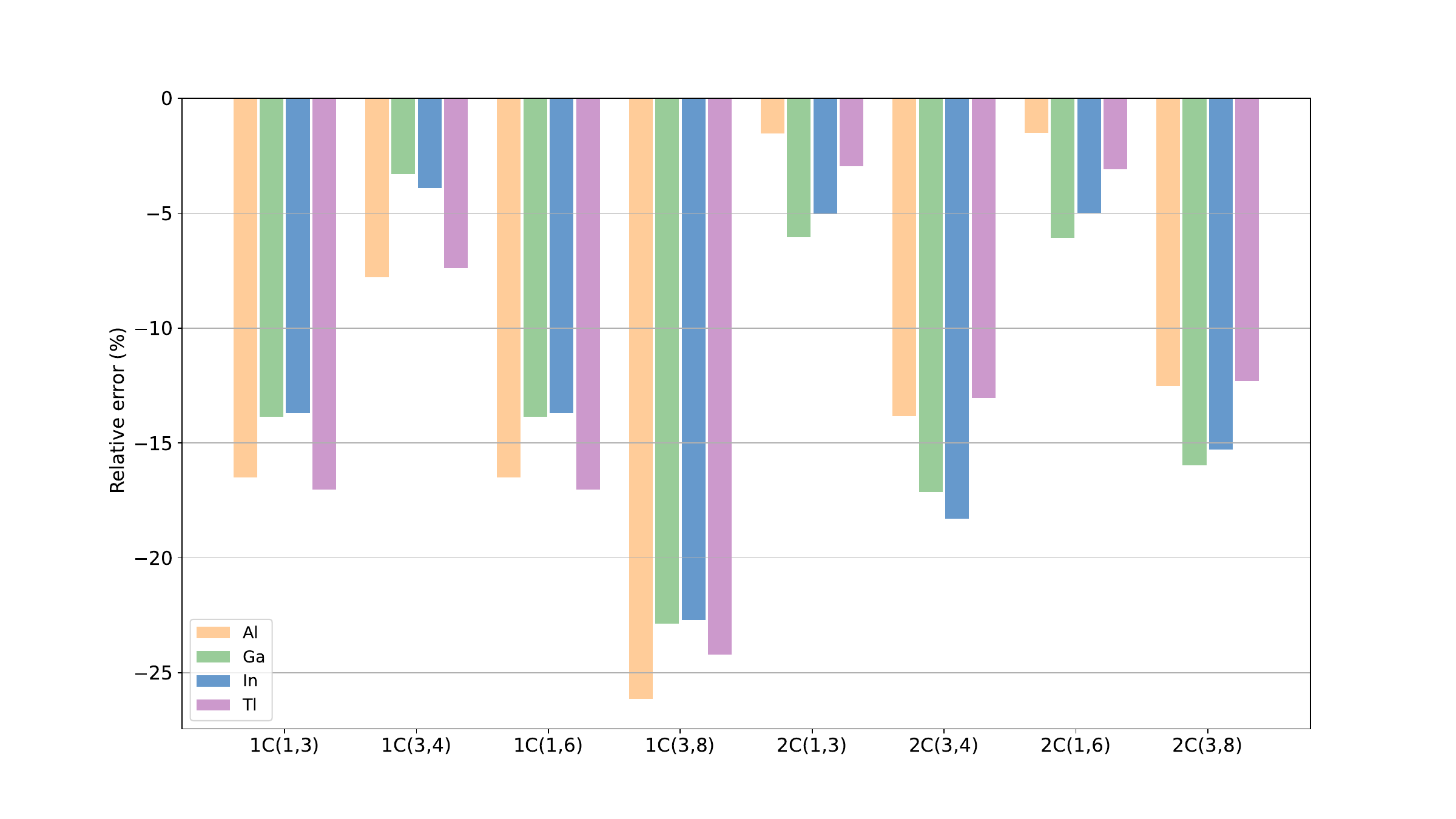}\\
                \end{tabular}
                \caption{The relative errors of SOSs of $^2P$ configuration of Al, Ga, In, and Tl computed by 1C- and 2C-CASSCF with four different active spaces.}
        \label{Rel_3}
\end{figure}

The SOSs of selected group 13 elements (Al, Ga, In, Tl) are summarized in Figure \ref{Rel_3}. The 1C-CASSCF method systematically underestimates the SOSs across all systems. The best performance is achieved with the CAS(3,4) active space, where relative errors range from 3.29\% (Ga) to 7.40\% (Tl), all below 8\%. However, expanding the active space to CAS(1,6) or CAS(3,8) does not improve accuracy. In fact, the errors increase significantly with the size of the active spaces, with CAS(3,8) yielding relative errors exceeding 22\% for all elements. In contrast, the 2C-CASSCF method achieves higher accuracy when only the $np^1$ electron is included in the active space, as in CAS(1,3) and CAS(1,6). 
These minimal active spaces yield remarkably low relative errors, down to 1.52\% for Al and 3.08\% for Tl, demonstrating excellent agreement with experimental data. In contrast, including the $ns^2$ electrons in the active space (as in CAS(3,4) and CAS(3,8)) results in larger underestimations, with the error increasing to 13.03\% for Tl. Notably, this trend parallels observations in group 14 elements, where the exclusion of $ns$ electrons in 2C-CASSCF improves the accuracy of SOSs of the $^3P$ configuration.

The SOSs of the $^3P$ states for chalcogen elements (S, Se, Te, Po) are presented in Tables S4 (2C-CASSCF) and S9 (1C-CASSCF) of the Supporting Information (SI). The $^3P$ configuration splits into three levels, $^3P_2$, $^3P_1$, and $^3P_0$, with $^3P_2$ being the ground state. 
For the $^3P_1$--$^3P_2$ splitting, the 2C-CASSCF method shows excellent agreement with experiment, with relative errors below 3.5\% across all elements and active spaces. 
In contrast, 1C-CASSCF results exhibit significantly larger deviations, particularly for heavier elements. The relative errors reach -11.1\% for Te in all calculations, indicating a systematic underestimation under strong SOC.

In chalcogen elements, the $^3P_1$--$^3P_0$ energy gap poses a greater challenge due to a level inversion along the series. Experimentally, $^3P_1$ is lower than $^3P_0$ for S and Se, but the order reverses at Te, where $^3P_0$ lies 44.2 cm$^{-1}$ below $^3P_1$. 
This crossover is not captured by 1C-CASSCF, which incorrectly predicts $^3P_1$ with a lower energy for Te. 
Even the 2C-CASSCF method fails to reproduce the correct ordering with smaller active spaces. Both CAS(6,4) and CAS(4,6) predict the $^3P_1$ state of Te as more stable. The result reported by Jenkins and coworkers using the two-component CASSCF method, with the CAS(6,4) active space, also predicts that $^3P_1$ is more stable than $^3P_0$.\cite{LixiaosongX2CCASSCF}
However, with the expanded CAS(6,8) active space, 2C-CASSCF successfully predicts the correct energy order for Te, with a $^3P_1$--$^3P_0$ splitting of $16.1~\mathrm{cm}^{-1}$, in qualitative agreement with experiment. 

\begin{figure}[!htp]
        \centering
                \begin{tabular}{c}
                    \includegraphics[width=1.0\textwidth]{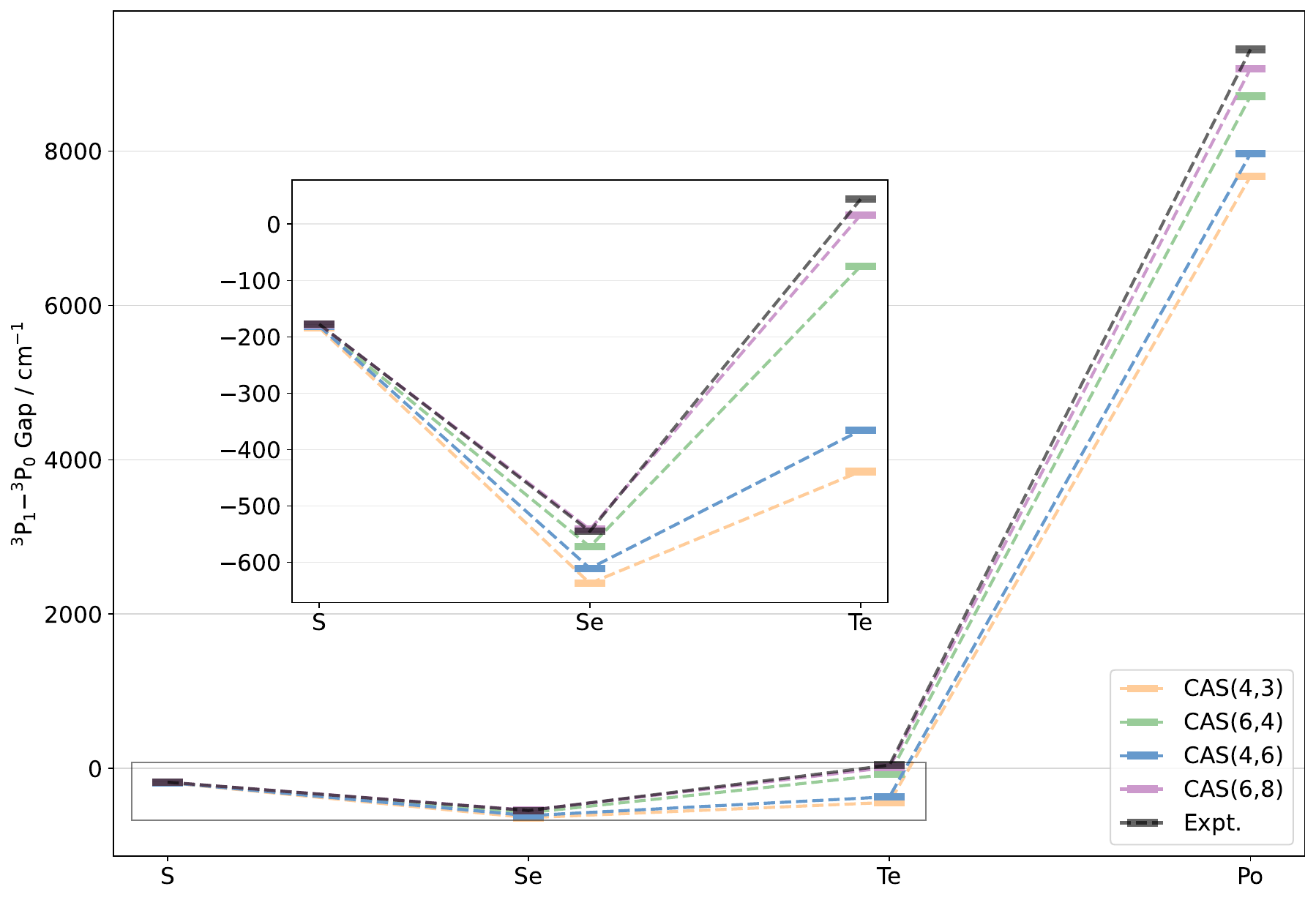}\\
                \end{tabular}
                \caption{The $^3P_1$--$^3P_0$ gaps (cm$^{-1}$) of S, Se, Te, and Po computed by 2C-CASSCF with four different active spaces.}
        \label{P1-P0}
\end{figure}

\begin{figure}[!htp]
        \centering
                \begin{tabular}{c}
                    \includegraphics[width=1.0\textwidth]{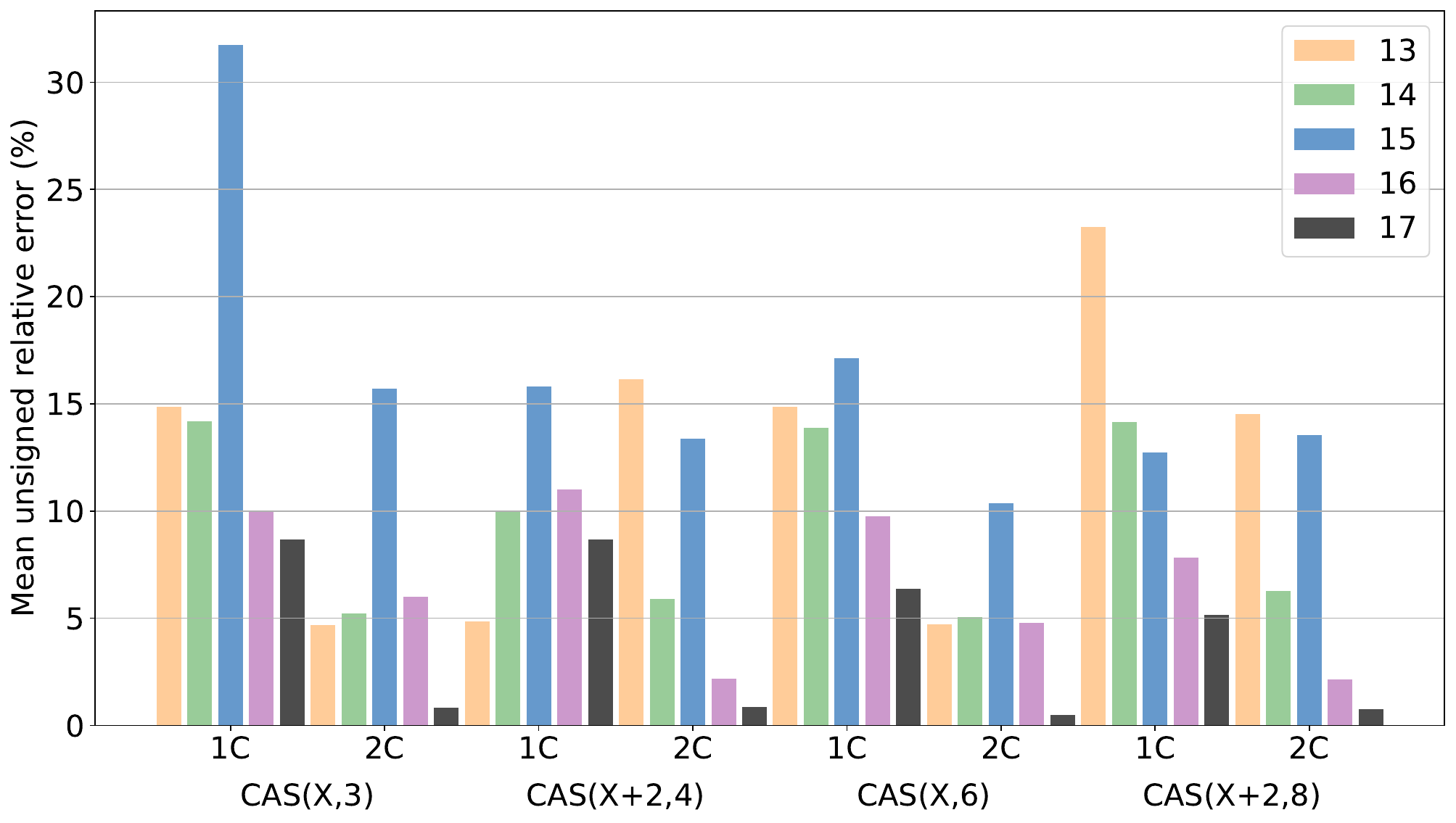}\\
                \end{tabular}
                \caption{The mean unsigned relative errors of SOSs of $p$-block elements from 4$^{th}$ to 6$^{th}$ row computed by 1C- and 2C-CASSCF. The experimental values are used as reference. The number of active electrons, X = [1, 5] for group 13 to 17.}
        \label{group}
\end{figure}

The SOSs of selected $p$-block elements computed using 1C- and 2C-CASSCF methods are given in the SI. Figure~\ref{group} presents the mean unsigned relative errors for both methods across groups 13--17 using four different active spaces, CAS(5,3), CAS(7,4), CAS(5,6), and CAS(7,8). Except for group 13 with CAS(3,4) and group 15 with the CAS(5,8) active space, the 2C-CASSCF method consistently yields significantly smaller relative errors compared to 1C-CASSCF, demonstrating superior accuracy in describing spin-orbit coupling effects. For 2C-CASSCF, active spaces including the $ns^2$ electrons, CAS(6,4) and CAS(6,8), deliver better performance for groups 16, with errors as low as 2.17\%. In contrast, for groups 13 and 14, active spaces containing only the $np$ electron, CAS(X,3) and CAS(X,6), yield the lowest errors (down to 4.69-5.07\% ), while inclusion of the $ns$ electrons (CAS(X+2,4), CAS(X+2,8)) increases the error, particularly for group 13. This indicates that, for these groups, restricting the active space to the valence $p$ orbitals helps avoid an imbalanced treatment of static correlation and spin–orbit coupling.

\subsection{HI and HAt}
\begin{figure}[!htp]
        \centering
                \begin{tabular}{cc}
         \includegraphics[width=0.5\textwidth]{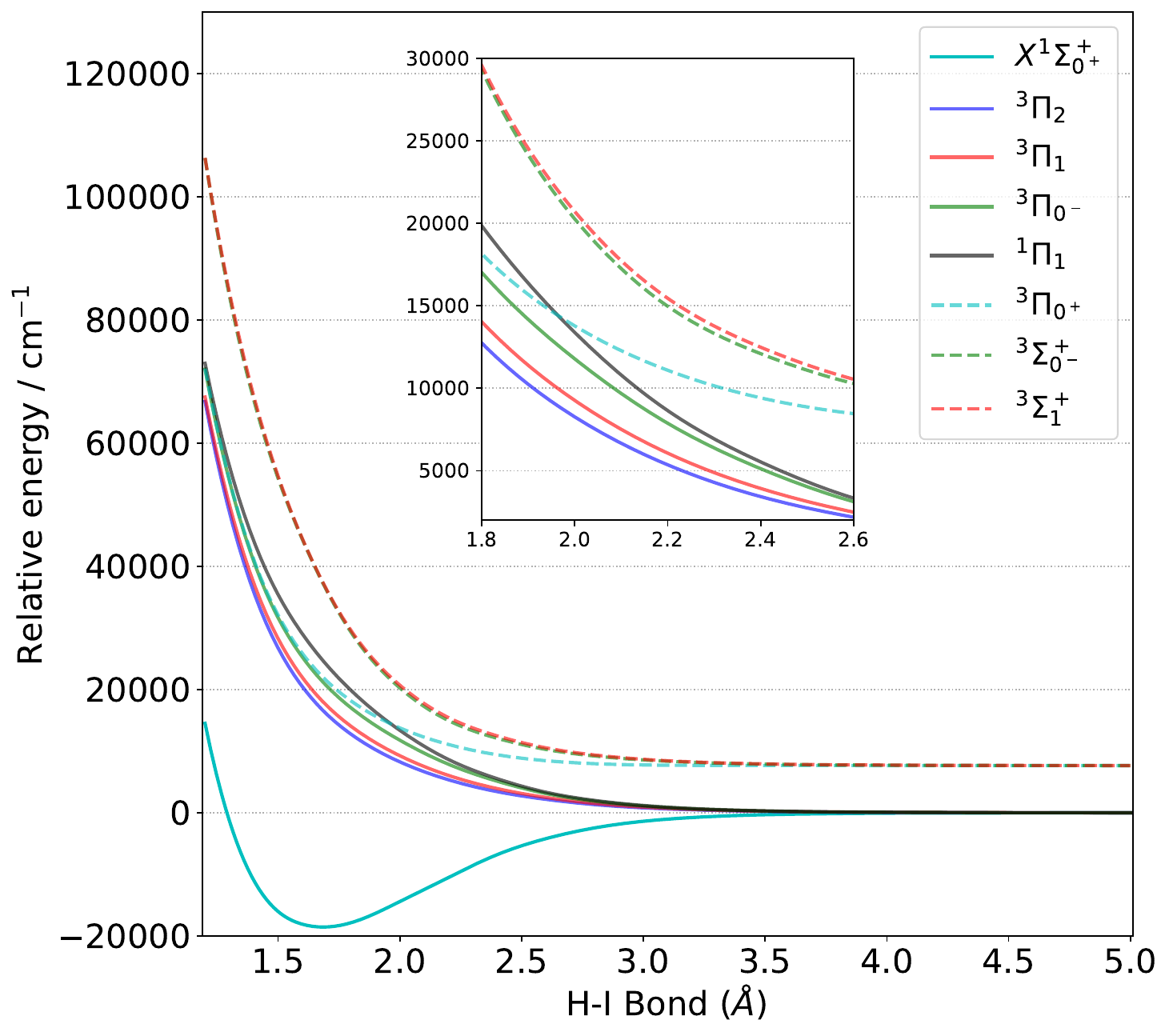} & \includegraphics[width=0.5\textwidth]{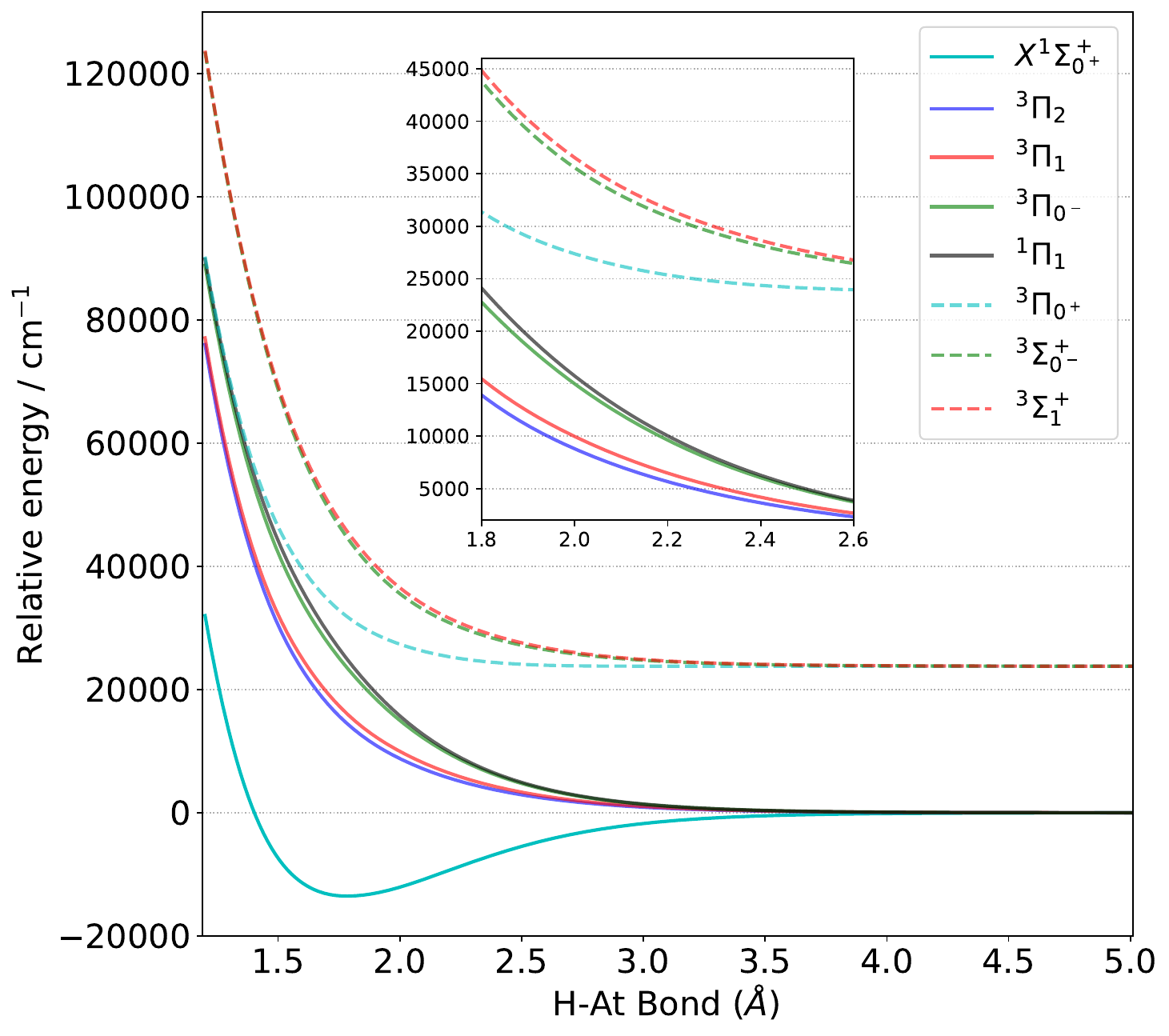}\\
                    (a) & (b)
                \end{tabular}
                \caption{The potential energy curves of ground and low-lying excited states of (a) HI and (b) HAt molecule, with respect to the dissociation limit of the ground state, $X ^1\Sigma^+_{0^+}$, computed using 2C-CAS(8,5).}
        \label{HIHAt}
\end{figure}

Besides the SOSs of $p$-block elements, the potential energy curves (PECs) of two diatomic molecules are studied using the 2C-CASSCF as well.
The ground ($X ^1\Sigma^+_{0^+}$) and low-lying excited state PECs of HI and HAt are computed using the 2C-CAS(8,5) active space and the dyallv3z basis set, considering eight spinor states.  The PECs of HI is well studied due to as  it is a popular candidate for probing excited electronic states Using vibrationally mediated photolysis\cite{camden}. As shown in Figure~\ref{HIHAt}, both molecules exhibit pronounced SOC effects, which split the three spin-free excited states ($^3\Pi$, $^1\Pi$, and $^3\Sigma^+$) into seven spinor states, labeled as $^3\Pi_{2}$, $^3\Pi_{1}$, $^3\Pi_{0^-}$, $^1\Pi_{1}$, $^3\Pi_{0^+}$, $^3\Sigma^+_{0^-}$, and $^3\Sigma^+_{1}$. For both HI and HAt, the excited state PECs are well separated from the ground state at equilibrium geometries. Upon bond elongation, these curves gradually converge to the dissociation limits $^2S_{1/2} + ^2P_{3/2}$ and $^2S_{1/2} + ^2P_{1/2}$. Compared to HI, HAt exhibits significantly stronger relativistic effects due to the presence of the astatine atom, resulting in larger energy splittings. Different from HI, the $^3\Pi_{0^+}$ state of HAt displays clear avoided crossings with the $^1\Pi_{1}$ state over a range of bond lengths ($R \approx 1.8$–$5.0$~\r{A}). These findings highlight the critical influence of relativistic effects on the photophysical behavior of heavy-element-containing molecules and emphasize the necessity of 2C approaches for accurately describing their excited-state dynamics. 

\section{Conclusions and Outlook}\label{Conclusion}

In the present work, we have implemented a two-component relativistic CASSCF method based on the perturbative Super-CI (Super-CIPT) approach for orbital optimization. The 2C-CASSCF method integrates SOC variationally at the self-consistent field level, which is able to describe relativistic effects and static correlation simultaneously. Benchmark calculations on SOSs of $p$-block atoms demonstrate that 2C-CASSCF significantly improves accuracy over traditional 1C-CASSCF with perturbative inclusion of spin-orbit coupling, even for the third row elements.

The use of the X2CAMF Hamiltonian, which incorporates the two-electron Gaunt or Breit term, is demonstrated to be essential for achieving quantitative accuracy, reducing relative errors to below 2\% for the SOSs of halogen elements. This confirmed the significance of including two-body interactions in high-precision theoretical predictions. The convergence behavior of Super-CIPT is also found to be stable and efficient for $p$-block elements. Although convergence is slightly slower compared to its spin-free counterpart, this is expected due to the added complexity of relativistic effects.

In the work, four different active spaces are used to compute SOSs of $p$-block elements. We found that the accuracy is slightly dependent on the choice of active space. For group 13 and 14 elements, compact active spaces containing only $np$ electrons yield the best performance. In contrast, group 16-17 elements require inclusion of $ns^2$ electrons to properly describe orbital relaxation and correlation under strong SOC. Nevertheless, to compute the SOSs of all $p$-block elements, the CAS(X,6) active spaces should be suggested. The scalar relatisvitic 

Due to the absence of dynamic correlation, the present 2C-CASSCF may not deliver accurate results for excited spinor states. Thus, one needs to incorporate dynamic correlation via second-order perturbation theory. These advances will enable accurate studies of excited states, and magnetic properties in heavy-element chemistry. Research is currently underway in this direction.

\section*{Acknowledgment}
This work was supported by the National Natural Science Foundation of China (Grant Nos. 22433001, and 22273052).
\appendix

\section*{Supporting Information}
See the Supporting Information for the spin-orbit splittings of selected $p$-block elements computed by 1C- and 2C-CASSCF methods.

\bibliography{iCI}

%
%\newpage
%For TOC only

%\includegraphics[width=\textwidth]{TOC.pdf}

\end{document}